\documentclass[a4paper,11pt]{article}

\usepackage[english]{babel}
\usepackage{lscape}
\usepackage{url}
\usepackage[toc,page]{appendix}
\usepackage{multicol}
\usepackage[squaren,Gray]{SIunits}
\usepackage{verbatim}
\usepackage{amsmath, amsthm,amsopn,amsfonts,amssymb,dsfont, esint, oldgerm,bm}
\usepackage{enumerate}
\usepackage{framed}
\usepackage{fullpage}
\usepackage{array}
\usepackage{graphicx}
\usepackage[font=small]{caption}
\usepackage[squaren,Gray]{SIunits}
\usepackage{longtable}
\usepackage{subcaption}
\usepackage{epsfig,wrapfig}
\usepackage{cite}


\author{David Petiteau*$^1$, Sebastien Guenneau$^1$, Michel Bellieud$^2$, Myriam Zerrad$^1$ \\and Claude Amra$^1$}
\title{Spectral efficiency of engineered thermal cloaks\\in the frequency regime}

\date{}
\begin{document}
\maketitle
\begin{center}
$^1$Aix Marseille Universit\'{e}, CNRS, Centrale Marseille, Institut Fresnel, UMR 7249, 13013 Marseille, France \\
$^2$LMGC, UMR-CNRS 5508, Universit\'{e} Montpellier II, 34095 Montpellier Cedex 5, France \\
david.petiteau@fresnel.fr
\end{center}
\abstract{We analyse basic thermal cloaks designed via different geometric transforms applied to thermal cloaking. We evaluate quantitatively the efficiency of these heterogeneous anisotropic thermal cloaks through the calculation of the standard deviation of the isotherms. The study addresses the frequency regime and we point out the cloak's spectral efficiencies. We find that all these cloaks have comparable efficiency irrespective of whether or not they have singular conductivity at their inner boundary. However, approximate cloaking with multi-layered cloak critically depends upon the homogenization algorithm and a large number of thin layers (at least fifty) is required to reduce substantially the standard deviation of the isotherms.}

\section*{Introduction}
In 2006, Pendry, Schurig and Smith \cite{pendry06} showed that upon a transformation applied to a certain region of space, one could obtain a phenomenon of optical invisibility via highly heterogeneous and anisotropic meta-materials whose permittivity and permeability directly result from the space transformation. Leonhardt proposed a similar route to cloaking based upon a conformal map preserving right angles, which makes the required meta-materials isotropic \cite{leonhardt06}, but this only works for a certain frequency range. Following these works, similar techniques  have been used to generalize these concepts to acoustics and mechanics. In 2012, some of us proposed to extend concepts of cloaking to diffusion processes,
such as heat fluxes \cite{guenneau12}, which led to first experimental demonstrations \cite{schittny13, han14, xu14, han14_2} in the transient regime. Interestingly, cloaking can be also applied to mass diffusion in a very similar manner \cite{guenneau13} (this may open a new range of biomedical and chemical engineering \cite{lunwu13} applications as well as light diffusion \cite{schittny14}).

This paper is focused on the analysis of the cloaking efficiency in the case of heat diffusion. Indeed numerous studies point out the solution of ideal cloaking but fewer evaluate in a quantitative way the resulting efficiency after the homogenization procedure. To reach this goal the efficiency is here characterized by  a root-mean square calculated on the line fluxes, a value which would be zero in the case of a homogeneous medium. Such characteristics allow us to compare different homogenized cloaks and emphasize their advantages and weaknesses. Moreover, rather than analyzing the efficiency at each time step, we prefer to perform a spectral analysis, on the basis of the classical time-frequency analogy. Hence we work in the frequency or time harmonic regime and investigate the spectral efficiency of various homogenized cloaks.
We first investigate both ideal
(singular and non singular) cloaks and approximate cloaks. Then, we improve earlier attempts to design multi-layered thermal cloaks \cite{guenneau12,han13} using the two-scale convergence method introduced in \cite{nguetseng89} and further developed by \cite{allaire92}. We show that our improved design reduces by at least one order of magnitude the standard deviation of thermal isovalues compared to earlier designs.  Potential applications are in thermal metamaterials and the management of heat flux.

\section*{Results}
	\subsection*{Quantitative analysis of different ideal heat cloak efficiencies}
		\subsubsection*{Space transformation and singularities}
Changes from polar coordinates $(r,\theta)$ to radially stretched polar coordinates $(r',\theta')=(f(r),\theta)$ applied to the heat equation (\ref{harmonic_heat_equation}) gives us the transformed conductivity and transformed product of heat capacity and density \cite{guenneau12}
\begin{equation}
\underline{\kappa}' = \kappa {\bf R}(\theta)\begin{pmatrix}
\kappa'_{rr} & 0 \\ 
0 & \kappa'_{\theta\theta}
\end{pmatrix}
{\bf R}(\theta)^T
\; , \;
\rho'c'(r')=\rho c\frac{g(r')\frac{dg}{dr'}(r')}{r'}
\label{eq1}
\end{equation}
where we have
\begin{equation}
\kappa'_{rr}(r') = \frac{g(r')}{r'\frac{dg}{dr'}(r')}\; , \;
\kappa'_{\theta\theta}(r') = \frac{1}{\kappa'_{rr}}
\label{kappa_r&kappa_theta}
\end{equation}
and $g$ is the reciprocal function of the monotonic function $f$, $\frac{dg}{dr'}$ is the derivative function of $g$, ${\bf R}$ is the rotation
matrix through an angle $\theta$ and ${\bf R}^T$ its
transpose. Note that these parameters hold for the transformed heat equation written in time or frequency domains (see Sec. \ref{time_harmonic_heat_equation}). Also, in the static limit, the product $\rho'c'$ does not appear in (\ref{eq1}), and one need only consider the anisotropic heterogeneous conductivity in (\ref{eq1}) to obtain the desired management of heat flux. Notice that a particularly appealing application of the transformed heat equation is thermal cloaking \cite{alu14}.
It can be more convenient to write relations (\ref{eq1}-\ref{kappa_r&kappa_theta}) versus the variable $r=g(r')$, with $\frac{dg}{dr'} =  \frac{1}{\frac{df}{dr}}$. The result is:
\begin{equation}
\kappa'_{rr} = \frac{r}{f(r)}\frac{df}{dr}(r)\; , \; \kappa'_{\theta\theta} = \frac{1}{\kappa'_{rr}}\; , \; \rho'c' = \rho c\frac{r}{f(r)\frac{df}{dr}(r)}
\label{general_cloak_parameters}
\end{equation}	
These last equations directly recall that singularities occur at $r' = f(r)_{|r=0}$, a radius where $\kappa'_{rr}$ is zero and $\kappa'_{\theta\theta}$ infinite, while $\rho'c'$ depends on the behaviour of $f$ around the origin. This is true whatever the transformation $f$ though $r=0$ does not necessarily concern the transformed region. Note in passing that a spherical (3D) diffusion cloak would have a different kind of singularity, with the radial component of the conductivity  vanishing at the inner boundary of the cloak, while the other two are constant \cite{guenneau13}.

		\subsubsection*{Specific transformations}
Now different $f$ functions can be used to obtain a simple design of ideal cloak by mapping a ring $\varepsilon R_1 \leq r \leq R_2$ (with $\varepsilon < \frac{R_2}{R_1}$) into another ring $R_1 \leq r' \leq R_2$. The following conditions must be fulfilled
\begin{equation}
f(\varepsilon R_1) = R_1\; , \; f(R_2) = R_2
\end{equation}
Depending on whether $\varepsilon < 1$ or $\varepsilon >1$, the $f$ transformation characterizes a contraction or dilatation, respectively.
A basic transformation may consist in a polynomial function $f(r) = \sum_{n=0}^N a_n r^n$.

Pendry's and Milton's cloaks \cite{pendry06, milton07} were based on linear ($N = 1$) and quadratic ($N = 2$) maps, respectively. Both also used $\varepsilon = 0$, which means that the whole disk $0 \leq r \leq R_2$ is transformed into the ring $R_1 \leq r' \leq R_2$.   
With $N = 1$ (Pendry), we obtain		
\begin{equation}
f(r) =  R_1 + r\frac{R_2-R_1}{R_2}
\label{Pendry_transform}
\end{equation}
which leads to
\begin{subequations}
\begin{align}
\kappa'_{rr} &= \frac{R_2-R_1}{R_2}\frac{r}{R_1+r\frac{R_2-R_1}{R_2}} = \frac{r'-R_1}{r'} \\
\kappa'_{\theta\theta} &= \frac{R_2}{R_2-R_1}\frac{R_1+r\frac{R_2-R_1}{R_2}}{r} = \frac{r'}{r'-R_1} \\
\rho'c' &= \rho c\frac{r}{\frac{R_2-R_1}{R_2}\left(R_1 + r\frac{R_2-R_1}{R_2}\right)} = \left(\frac{R_2}{R_2-R_1}\right)^2\frac{r'-R_1}{r'}
\end{align}
\label{parametres_Pendry}
\end{subequations}
with $\kappa'_{rr}$ varying in the range $\left[0,\frac{R_2-R_1}{R_2}\right]$, $\kappa'_{\theta\theta}$ in the range $\left[\frac{R_2}{R_2-R_1}, +\infty\right[$ and $\rho'c'$ in the range $\left[0, \frac{R_2}{R_2-R_1}\right]$ when $r' \in \left[R_1, R_2\right]$.
We note the discontinuity of $\kappa'_{rr}$, $\kappa'_{\theta\theta}$ and $\rho'c'$ with the parameters of the surrounding medium at the interface $r'=R_2$.

With $N = 2$, one of the $a_n$ coefficients (namely $a_1$ or $a_2$) can be arbitrarily chosen. To remove such an indetermination Milton proposed to ensure the continuity of the transformed conductivity $\underline{\kappa}'=diag(\kappa'_{rr}, \kappa'_{\theta\theta})$ and $\rho'c'$ product at $r' = R_2$. We can see from (\ref{general_cloak_parameters}) that this continuity at $r' = R_2$ amounts to ensure that $\frac{df}{dr}_{|r=R_2} = 1$ giving the following expressions on the $a_n$ coefficients
\begin{equation}
a_2 = \frac{R_1}{R_2^2}\;,\;a_1 = 1-2\frac{R_1}{R_2}\;\text{and}\;a_0=R_1
\end{equation}
This gives us
\begin{subequations}
\begin{align}
\kappa'_{rr} &=\frac{r(2a_2r+a_1)}{a_2r^2+a_1r+a_0} = \frac{a_1^2+4a_2\left(r'-a_0\right)-a_1\sqrt{a_1^2+4a_2\left(r'-a_0\right)}}{2r'a_2} \\
\rho'c' &= \rho c\frac{r}{(2a_2r+a_1)(a_2r^2+a_1r+a_0)}=\rho c\frac{\sqrt{a_1^2+4a_2\left(r'-a_0\right)}-a_1}{2r'a_2\sqrt{a_1^2+4a_2\left(r'-a_0\right)}}
\end{align}
\end{subequations}
and $\kappa'_{\theta\theta} = \frac{1}{\kappa'_{rr}}$. $\kappa'_{rr}$ and $\kappa'_{\theta\theta}$ both vary in the range $\left[0,1\right]$ and we have $\rho'c' \in \left[0,\rho c\right]$ when $r' \in \left[R_1, R_2 \right]$. Then, we clearly see that the continuity of $\kappa'_{rr}$, $\kappa'_{\theta\theta}$ and $\rho'c'$ is ensured at the interface $r'=R_2$ with the surrounding media.

The case $\varepsilon \neq 0$ avoids the singularity at $r' = R_1$. With $N = 1$, Kohn \cite{kohn08} obtained $a_1 = \frac{R_2-R_1}{R_2-\varepsilon R_1}> 0$ and $a_0 = (1-\varepsilon)\frac{R_1R_2}{R_2-\varepsilon R_1} \neq 0$ leading to
\begin{subequations}
\begin{align}
\kappa'_{rr} &= \frac{a_1r}{a_1r+a_0} = \frac{r'-a_0}{r'} \\
\kappa'_{\theta\theta} &= \frac{a_1r+a_0}{a_1r} = \frac{r'}{r'-a_0} \\
\rho'c' &= \rho c\frac{r}{a_1\left(a_1r+a_0\right)} = \frac{r'-a_0}{a_1^2r'}
\end{align}
\end{subequations}
with $\kappa'_{rr}$ varying in the range $\left[\varepsilon\frac{R_2-R_1}{R_2-\varepsilon R_1}, \frac{R_2-R_1}{R_2-\varepsilon R_1}\right]$, $\kappa'_{\theta\theta}$ in $\left[\frac{R_2-\varepsilon R_1}{R_2-R_1}, \frac{R_2-\varepsilon R_1}{\varepsilon(R_2-R_1)}\right]$ and $\rho'c'$ in $\left[\rho c\varepsilon\frac{R_2^2}{(R_2-\varepsilon R_1)(R_2-R_1)}, \rho c\frac{R_2-\varepsilon R_1}{R_2-R_1}\right]$ when $r' \in \left[R_1, R_2\right]$. If $\varepsilon \to 0$, we retrieve the expressions from (\ref{parametres_Pendry}).
  
\begin{figure}[ht!]
\centering
\includegraphics[width=\textwidth]{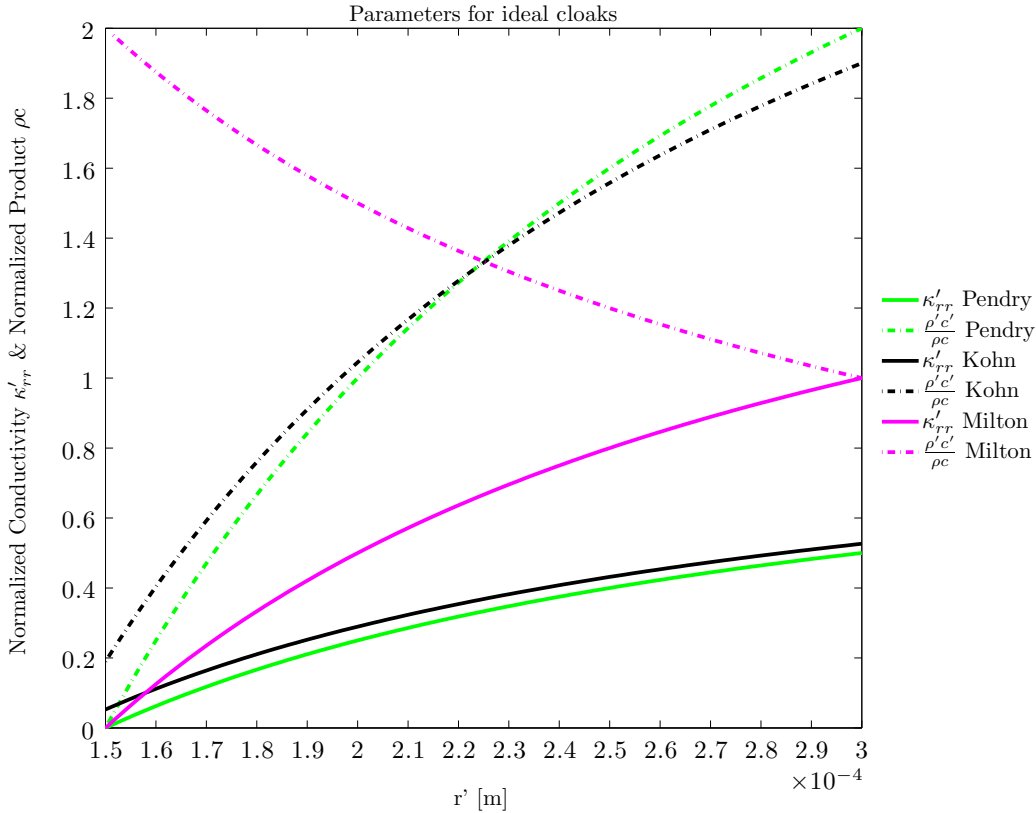}
\caption{\label{parameters_ideal_cloaks} Normalized conductivity $\kappa'_{rr}$ and product $\rho c$ parameters used for Pendry (green), Kohn (black) and Milton (magenta) cloaks as a function of $r'$ inside the cloak. The inner boundary of the cloak is $R_1 = 0.15$ mm and the outer boundary is $R_2 = 2R_1 = 0.3$ mm. We used here $\varepsilon = 0.1$ for Kohn's cloaks.}
\end{figure}

A summary of the thermal parameters' behaviour is given on Fig. \ref{parameters_ideal_cloaks} for different cloaks. Notice that with a general polynomial transformation the coefficients would be written as $a_n = a_n(\varepsilon,N)$. Other transformations could be used. With an exponential function we obtain $f(r) = R_1 \exp(-\mu\varepsilon R_1)\left(\frac{R_2}{R_1}\right)^{\frac{r}{R_2}}$.

		\subsubsection*{Implementation and efficiency of ideal cloaks}
We study the three cloaks introduced above using the commercial finite element package COMSOL Multiphysics \textregistered. Our two-dimensional simulation consists of the resolution of the time-harmonic heat equation (see Sec. \ref{time_harmonic_heat_equation}) in a $5$ mm by $0.8$ mm computational domain containing the studied cloak.  The medium surrounding the cloak corresponding to the region $r' \geq R_2$ is made of Titanium of thermal conductivity $\kappa_{Ti} = 22\;\text{W}.\text{m}^{-1}.\text{K}^{-1}$ and product of density by heat capacity $\rho c_{Ti} = 2.35\times 10^6\;\text{J}.\text{kg}^{-1}.\text{K}^{-1}$ giving a thermal diffusivity $a_{Ti} \approx 1.10^{-5}\;\text{m}^2.\text{s}^{-1}$. 
Then, the initial ring of thickness $R_2-\varepsilon R_1$ is mapped onto the ring of thickness $R_2-R_1$. This latter ring corresponds to the invisibility cloak whose parameters are shown on Fig. \ref{parameters_ideal_cloaks} depending on which transformation we choose (Pendry, Milton or Kohn). The effect of the cloak within $(R_1, R_2)$ is to straighten the isotherms for for $r'>R_2$ as if we were in an homogeneous medium. The inner and outer radii of the cloak are chosen to be respectively $R_1 = 0.15$ mm and $R_2 = 0.3$ mm.
Finally, the region $0 \leq r' \leq R_1$ (mapped from the initial region $0 \leq r \leq \varepsilon R_1$) contains the cloaked object. It is made of polyvinyl chloride (PVC) of thermal conductivity $\kappa_{PVC} = 0.15 \;\text{W}.\text{m}^{-1}.\text{K}^{-1}$ and product of density by heat capacity $\rho c_{PVC} = 1.33\times 10^6\;\text{J}.\text{kg}^{-1}.\text{K}^{-1}$ of thermal diffusivity $a_{PVC} \approx 1.10^{-7}\;\text{m}^2.\text{s}^{-1}$. 

Notice that the $5$ mm length of our computational domain has been chosen in such a way that it is in accordance with the thermal diffusion length $D$ in Titanium at angular frequency $\omega = 1$ rad.s$^{-1}$, $D=\sqrt{\frac{2a_{Ti}}{\omega}}= 4.4$ mm.
Regarding the boundary conditions, we chose to set the temperature to $T_0=1\;\text{K}$ on the left boundary of the computational domain and $T_f=0\;\text{K}$ on the right boundary of the computational domain so that heat diffuses from left to right (namely from $x=-2.5\;\text{mm}$ to $x=2.5\;\text{mm}$). As for the upper and lower boundaries of the computational domain, we chose Neumann (perfect insulating) conditions $\frac{\partial T}{\partial n} = 0$.

\begin{figure}[ht!]
\centering
\begin{subfigure}[c]{0.4\textwidth}
	\includegraphics[width=\textwidth]{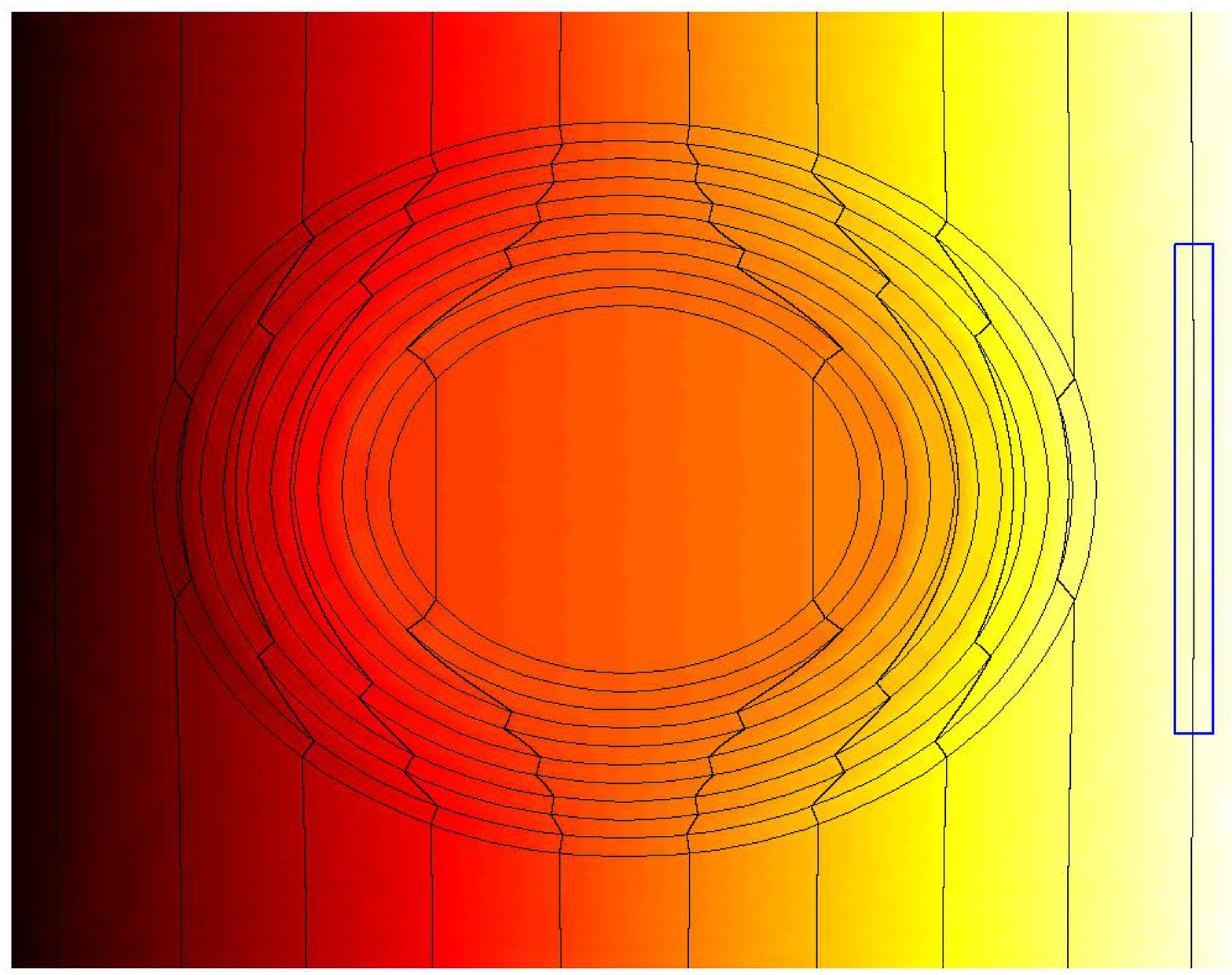}
\end{subfigure}
\begin{subfigure}[c]{0.4\textwidth}
	\includegraphics[width=\textwidth]{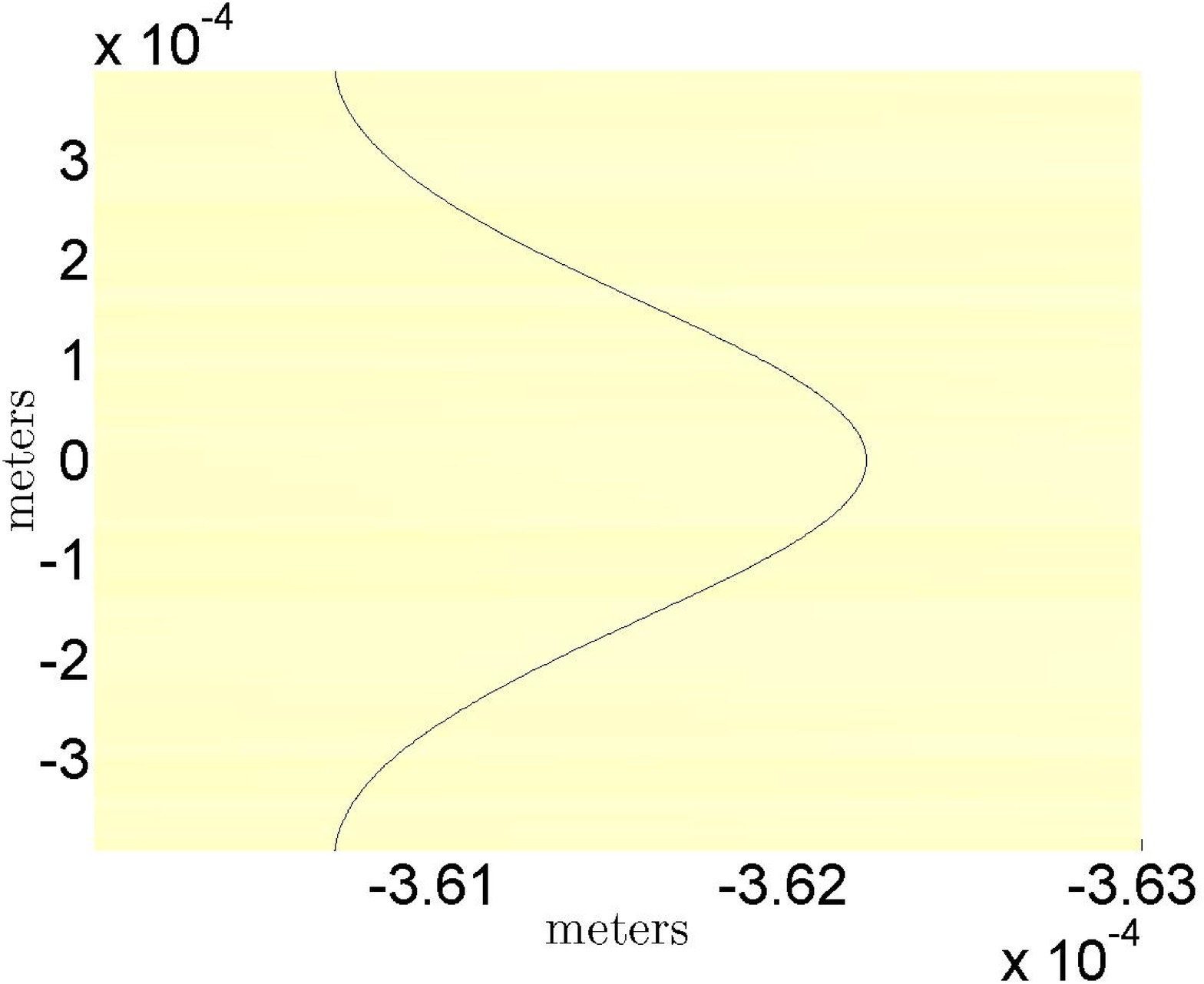}
\end{subfigure}
\caption{\label{cape_iso} Map of a thermal cloak in the static limit (left) with a close-up on the isotherms in its vicinity (right): The cloak's inner boundary is $R_1=0.15$ mm, and its outer boundary is $R_2=0.3$ mm.}
\end{figure}

Now the main idea to determine quantitatively the efficiency of a cloak is to evaluate the standard deviation of the isotherms in the vicinity of the cloak. A theoretical perfect cloak would give a zero standard deviation for all the isotherms outside the cloak as if the medium was completely homogeneous, which is clearly not the case in Fig. \ref{cape_iso}.
  
Thus, by evaluating the standard deviation of the isotherms (See Sec. \ref{standard_deviation_isotherms}) for different cloaks, we can quantitatively determine the efficiency of a cloak. All isotherms are calculated before and after the outer radius of the cloak in the incident diffusion direction and we define the standard deviation ratio
\begin{equation}
standard\;deviation\;ratio = \frac{std(bare\;object)}{std(studied\;cloak)}
\end{equation}
where $std(bare\;object)$ denotes the standard deviation of the isotherms when heat diffuses through the bare object to be cloaked and $std(studied\;cloak)$ is the standard deviation of the isotherms when the ideal cloak under study is around the object. An efficient cloak means that $std(studied\;cloak)$ is close to $0$. Therefore the $standard\;deviation\;ratio$ increases with increasing efficiencies.

\begin{figure}[ht!]
\centering
\includegraphics[width=\textwidth]{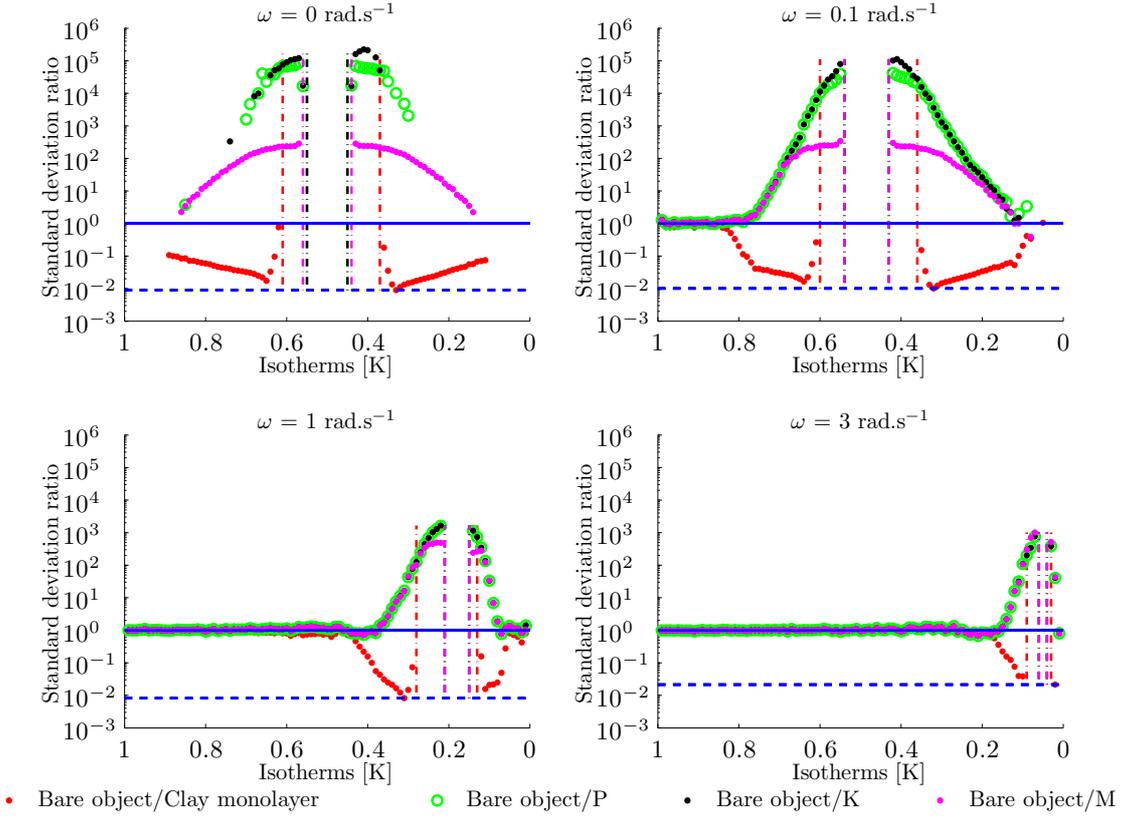}
\caption{Standard deviation ratio of a clay homogeneous monolayer cloak (red), Pendry's cloak (green), Kohn's cloak (black) and Milton's cloak (magenta) at angular frequencies $\omega = 0,\;0.1,\;1 \;\text{and}\; 3\;\text{rad}.\text{s}^{-1}$. The source is on the left so that heat diffuses from left to right. The vertical lines are related to the position of the insulating layer and cloaks.  The (z) vertical axis gives the ratio of the deviation on a logarithmic scale, while the (x) horizontal axis is for the temperature of the isotherms. Note that this temperature scale is directly connected to the x length coordinate. For the static regime for instance the temperature is $T = \frac{T_0+T_f}{2} = 0.5$ K at half of the total x length. This is no longer the case at increasing frequencies, which explains the shift of the curves with frequency $\omega$. The horizontal dashed blue line indicates the height of minimum of the standard deviation ratio among all the studied cloaks at frequency $\omega$.}
\label{rapport_capesparfaites}
\end{figure}

In Fig. \ref{rapport_capesparfaites}, we compare the standard deviation ratio of the isotherms resulting from a single insulating homogeneous layer made of clay of thermal conductivity $\kappa_{clay} = 1.28\;\text{W}.\text{m}^{-1}.\text{K}^{-1}$, product of density by heat capacity $\rho c_{clay} = 1.28\times 10^6\;\text{J}.\text{kg}^{-1}.\text{K}^{-1}$ (thermal diffusivity $a_{clay} \approx 1.10^{-6}\;\text{m}^2.\text{s}^{-1}$) in red, a perfect cloak (Pendry) calculated with $\varepsilon = 0$ and $N=1$ (green), a perfect cloak (Kohn) calculated with $\varepsilon \neq 0$  and $N=1$ (black), and a perfect cloak based on the Milton transformation (magenta) calculated with $\varepsilon = 0$ and $N = 2$.
Moreover calculation is performed at different frequencies, in a range appropriate with the thermal diffusion length and the deviation is plotted versus the average diffusion direction.

\begin{figure}[ht!]
\centering
\includegraphics[width=0.85\textwidth, keepaspectratio]{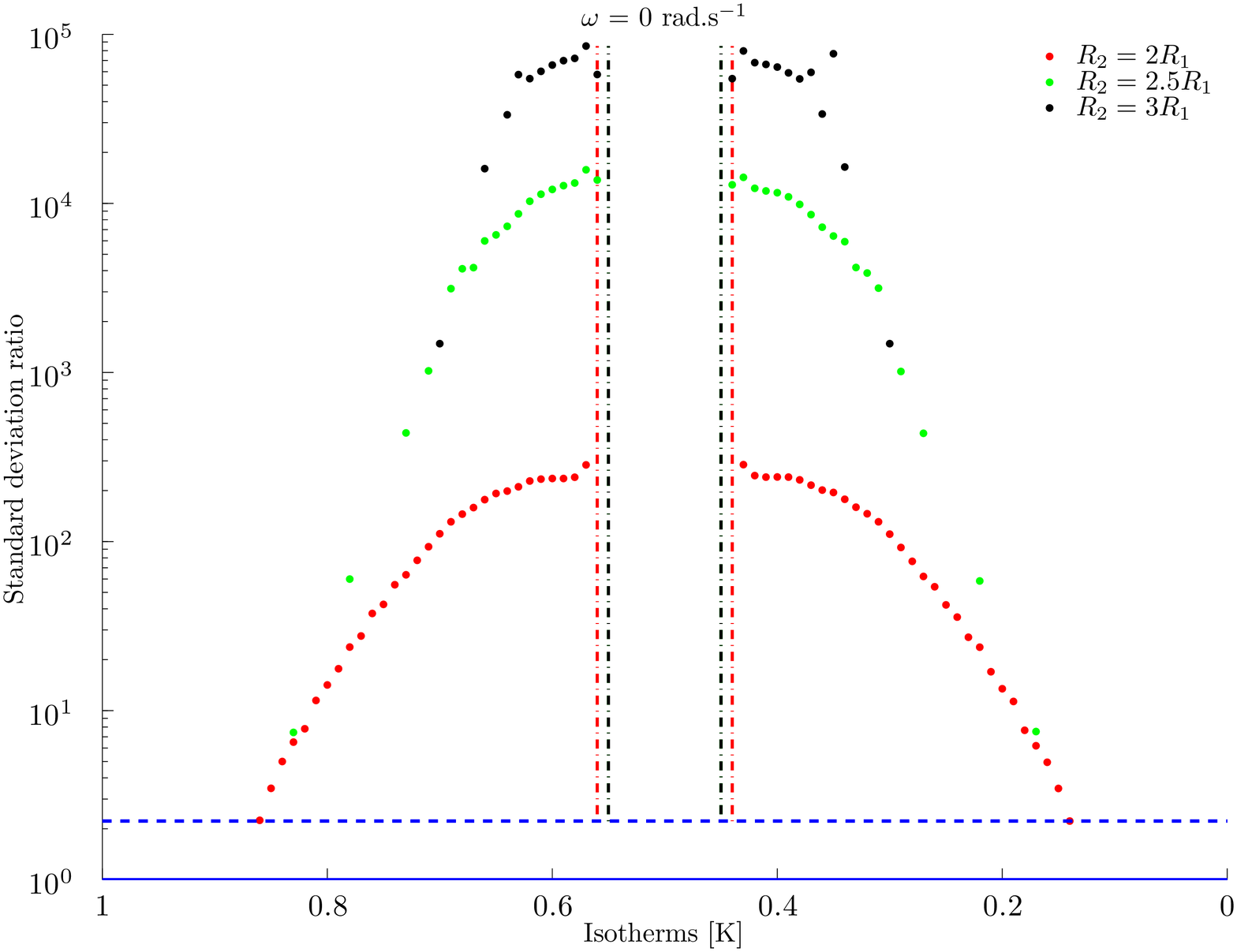}
\caption{Standard deviation ratio of Milton's cloaks at angular frequencies $\omega = 0\;\text{rad}.\text{s}^{-1}$ with $R_2 = 2R_1$ (red), $R_2=2.5R_1$ (green) and $R_2 = 3R_1$ (black). Note that the black curve is similar to green (Pendry's cloak) and black (Kohn's cloak) curves in the upper left panel of Fig. \ref{rapport_capesparfaites}. The horizontal dashed blue line indicates the height of minimum of the standard deviation ratio among the studied Milton's cloaks.}
\label{rapport_capesparfaites_R2=3R1}
\end{figure}

In Fig. \ref{rapport_capesparfaites}, the horizontal blue line at altitude $1$ is for the bare object, which is a hallmark of neutral efficiency. We first notice that the clay monolayer cloak is below this curve, which means that the deviation has been increased. On the other hand, as expected we observe in the vicinity of the cloak that the ideal Pendry and Kohn cloaks reduce the deviation by a factor of nearly $10^4$, which emphasizes a great efficiency. The Milton cloak appears to be less efficient, but this is due to the fact that Milton's transformation is defined for a whole range of radii $R_2 \geq 2R_1$ while we considered $R_2 = 2R_1$ in our case. The Milton cloak is therefore used in a critical regime explaining its lower performances. A design of Milton's cloak with $R_2 > 2R_1$ reaches identical efficiencies to that of Pendry's and Kohn's cloaks as we can see in Fig. \ref{rapport_capesparfaites_R2=3R1}.

	\subsection*{Towards homogenized multi-layered heat cloaks}
   
\begin{figure}[ht!]
\centering
\includegraphics[width=\textwidth]{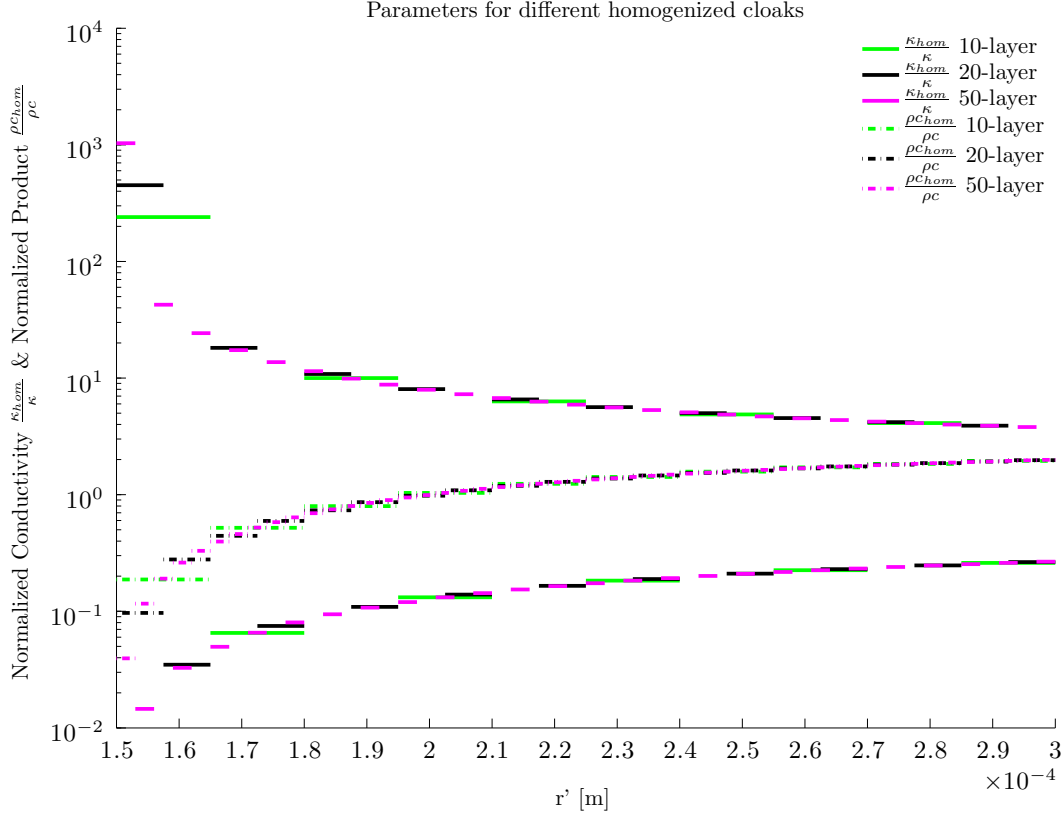}
\caption{\label{parameters_homogenized} Normalized conductivity $\frac{\kappa_{hom}}{\kappa}$and product $\frac{\rho c_{hom}}{\rho c}$ parameters used for the Pendry's 10-layer (green), 20-layer (black) and 50-layer (magenta) homogenized cloaks as a function of $r'$ inside the cloak (See Table \ref{table_10_layer} for explicit values used for the 10-layer homogenized cloak). $\kappa_{hom}$ denotes either the radial average $<\kappa_1>$ or $<\kappa_2>$ depending on the considered radial element. $\rho c_{hom}$ denotes the radial average $<\rho'c'>$ on the considered radial element.}
\end{figure}

We now want to build thermal cloaks that have similar properties and efficiencies as those of perfect cloaks. Thus we need a tool to design thermal meta-materials that will give the desired properties. For this purpose, we make use of the homogenization process (see Sec. \ref{homogenization_process}).
We start from an alternation of concentric homogeneous thin layers with respective conductivities and product of density by heat capacity $(\kappa_1, \rho_1c_1)$ and $(\kappa_2, \rho_2c_2)$ which might depend on $r'$. If the number of concentric layers gets large enough (and their thicknesses vanish), this alternation will behave as the required anisotropic inhomogeneous conductivity $\kappa'$  and inhomogeneous product $\rho'c'$ under certain conditions on $\kappa_1(r')$, $\kappa_2(r')$, $\rho_1c_1(r')$ and $\rho_2c_2(r')$:
\begin{equation}
\left\{\begin{array}{l l l}
\frac{2\kappa_1(r')\kappa_2(r')}{\kappa_1(r')+\kappa_2(r')} = \kappa_{rr}' = \frac{g(r')}{r'g'(r')} \\
\\
\frac{1}{2}(\kappa_1(r') + \kappa_2(r')) = \kappa_{\theta\theta}' = \frac{r'g'(r')}{g(r')}
\\ \\
\frac{1}{2}(\rho_1 c_1(r') + \rho_2 c_2(r')) = \rho'c' = \rho c\frac{g(r')g'(r')}{r'}
\end{array} \right.
\label{real_homog_system}
\end{equation}
We now consider the linear transformation $r' = a_1 r + a_0$ with $\varepsilon = 0$. Therefore, we want our homogenized cloak to behave like Pendry's cloak.
Solving for $\kappa_1$ and $\kappa_2$ provides us with the parameters to use in the concentric multi-layered cloak:
\begin{equation}
\begin{split}
\kappa_1(r') &= \frac{r'+\sqrt{R_1\left(2r'-R_1\right)}}{r'-R_1} \\
\kappa_2(r') &= \frac{r'-\sqrt{R_1\left(2r'-R_1\right)}}{r'-R_1}
\end{split}
\end{equation}
Then, we approximate $\kappa_1(r')$ and $\kappa_2(r')$ by dividing radially the cloak into $N$ elements and by taking the mean of $\kappa_1(r')$ and $\kappa_2(r')$ on each element. Also, we will assume $\rho_1c_1(r') = \rho_2c_2(r') = \rho'c'$. The Figure \ref{parameters_homogenized} shows the
evolution of the conductivity and product $\rho c$ parameters as one goes from the inner boundary of the cloak ($R_1 = 1.5$ mm) to the outer boundary of the cloak ($R_2 = 3$ mm) for Pendry's 10-layer, 20-layer and 50-layer homogenized cloaks.

\begin{figure}[ht!]
\centering
\includegraphics[width=\textwidth]{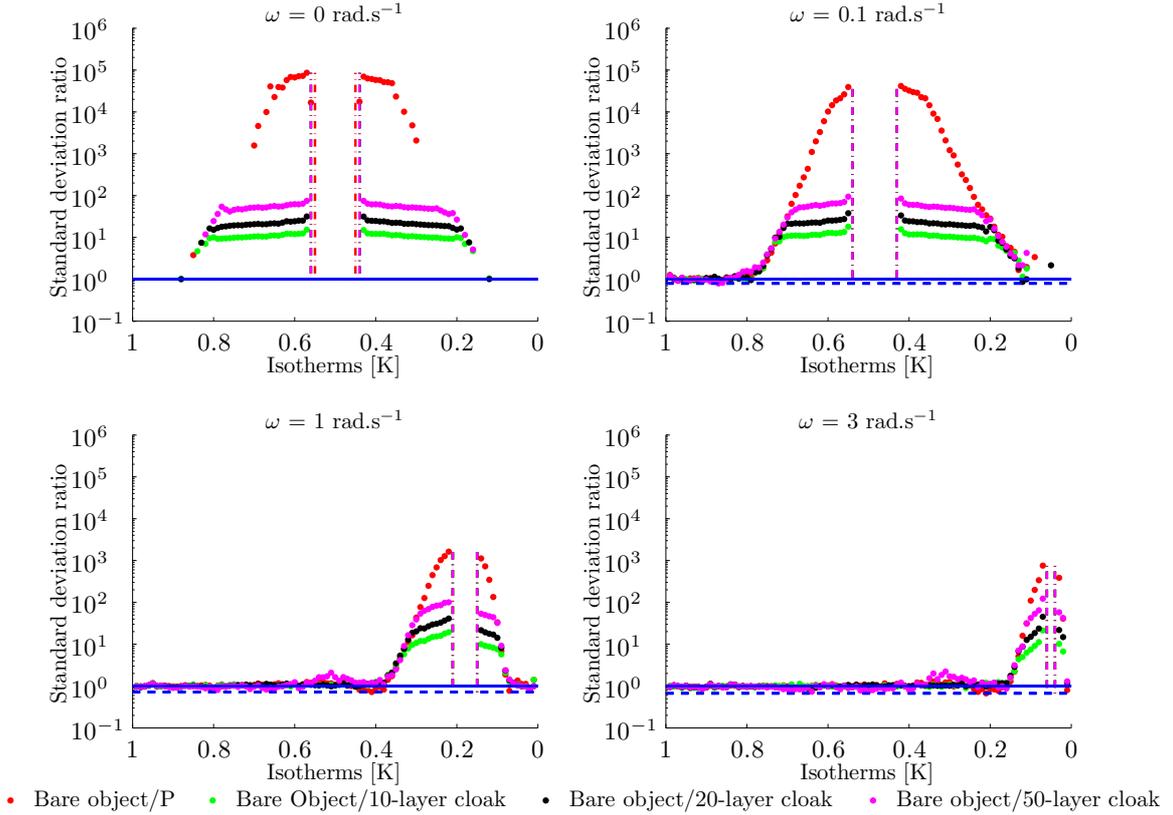}
\caption{\label{stdratio_Pendry_capeshomog} Standard deviation ratio of Pendry's perfect cloak (red), a 10-layer homogenized cloak (green), a 20-layer homogenized cloak (black) and a 50-layer homogenized cloak (magenta) at angular frequencies $\omega = 0,\;0.1,\;1 \;\text{and}\; 3\;\text{rad}.\text{s}^{-1}$. The source is on the left so that heat diffuses from left to right. The vertical lines indicate the position of the cloaks.  The (z) vertical axis gives the ratio of the deviation on a logarithmic scale, while the (x) horizontal axis is for the temperature of the isotherms. The horizontal dashed blue line indicates the height of minimum of the standard deviation ratio among all the studied cloaks at angular frequency $\omega$.}
\end{figure}

In Fig. \ref{stdratio_Pendry_capeshomog} we compare the $standard\;deviation\;ratio$ of Pendry's cloak (red), the 10-layer homogenized cloak (green), the 20-layer homogenized cloak (black) and the 50-layer homogenized cloak (magenta) introduced above. We can see that the homogenization process gives good results as the standard deviation ratio of the homogenized cloaks increases with the number of layers used, rising from $10^1$ for the 10-layer cloak in the vicinity of the cloak to $10^2$ for the 50-layer cloak. However, the goal is still far off since we would like to reach a standard deviation ratio of $10^4$ to ascertain our homogenized cloaks mimic Pendry's cloak. But we expect this gap between perfect and homogenized cloaks to depend merely on the number of layers used and the asymptotic behaviour of $\kappa_1$ around $r' = R_1$. This point is discussed in the next section.

\section*{Discussion}
\begin{figure}[ht!]
\includegraphics[width=\textwidth]{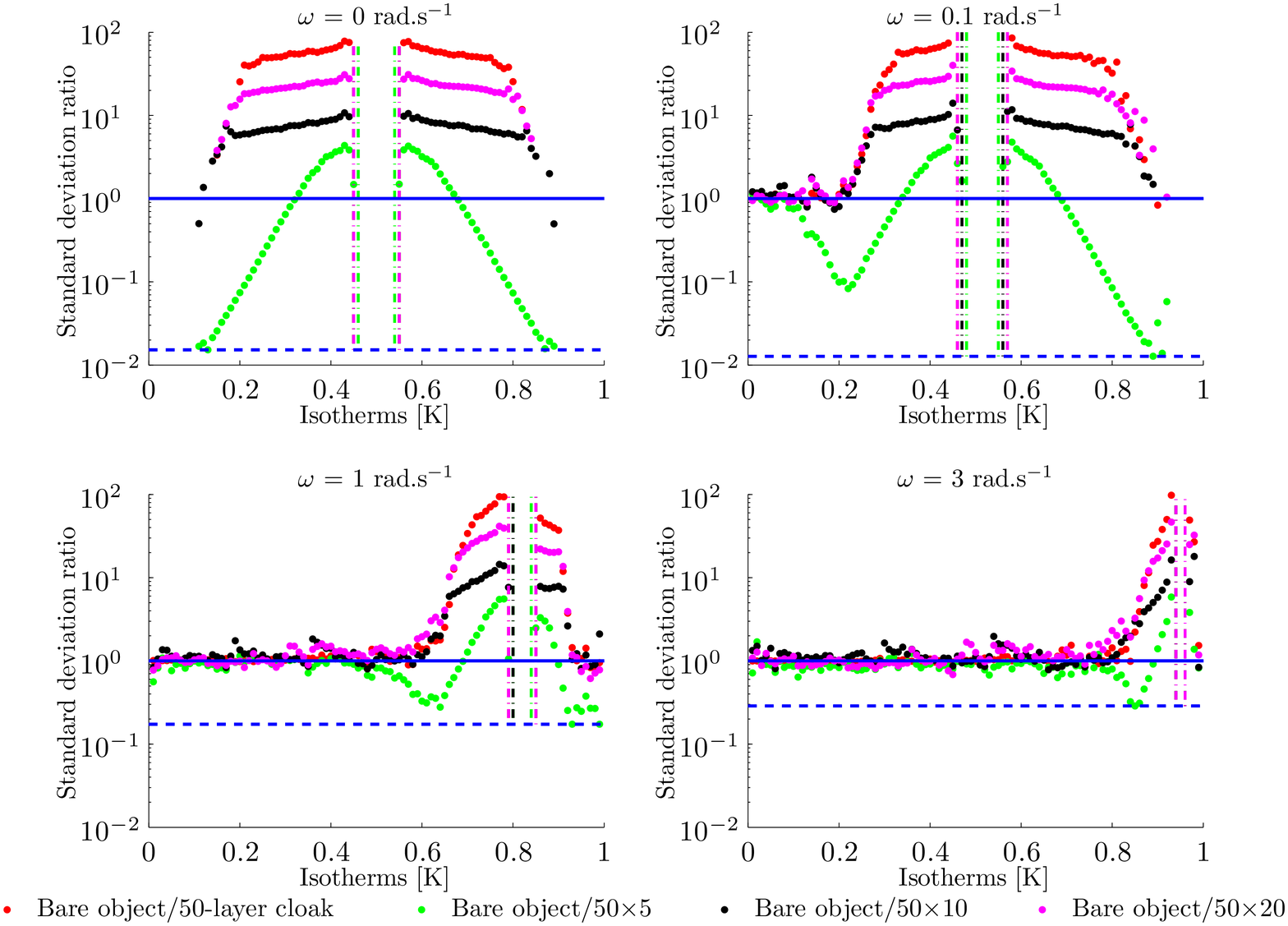}
\caption{\label{stdratio_capeshomog_capesdiscretes} Standard deviation ratio of a 50-layer homogenized isotropic cloak (red), a 50$\times$5 discretized anisotropic cloak (green), 50$\times$10 discretized anisotropic cloak (black) and a 50$\times$20 discretized anisotropic cloak (magenta) at angular frequencies $\omega = 0,\;0.1,\;1 \;\text{and}\; 3\;\text{rad}.\text{s}^{-1}$. The source is on the left so that heat diffuses from left to right. The vertical lines indicate the position of the cloaks.  The (z) vertical axis gives the ratio of the deviation on a logarithmic scale, while the (x) horizontal axis is for the temperature of the isotherms. The horizontal dashed blue line indicates the height of minimum of the standard deviation ratio among all the studied cloaks at frequency $\omega$.}
\end{figure}

To emphasize the efficiency of the homogenized cloak, we studied a discretized version of the perfect anisotropic Pendry's cloak in (\ref{Pendry_transform}) with $\varepsilon=0$. We divide the cloak  into $N\times M$ elements ($N$ elements in the radial direction and $M$ elements in the azimuthal directions). In each element, we take the average of the conductivity $\underline{\kappa}'$ so that the resulting conductivity in this element can be written
\begin{equation}
\begin{split}
<\underline{\kappa}'>_{\substack{r_i \rightarrow r_{i+1}\\ \theta_i \rightarrow \theta_{i+1}}} &= \frac{2}{(r_{i+1}^2-r_i^2)(\theta_{i+1}-\theta_i)}\int_{r_i}^{r_{i+1}}\int_{\theta_i}^{\theta_{i+1}} \underline{\kappa}' \;r\mathrm{d}r\mathrm{d}\theta\\
&= \begin{pmatrix}
<\kappa_{11}> & <\kappa_{12}> \\
<\kappa_{21}> & <\kappa_{22}>
\end{pmatrix}
\end{split}
\end{equation}
where $r_i$ and $r_{i+1}$ are the radial boundaries of the element, and $\theta_i$ and $\theta_{i+1}$ are the azimuthal boundaries of the element. Thus, we obtain a piecewise homogeneous anisotropic cloak. If $N$ and $M$ go to infinity we expect to retrieve the original anisotropic heterogeneous cloak.
In Fig. \ref{stdratio_capeshomog_capesdiscretes}, we compare the $standard\;deviation\;ratio$ of a 50-layer homogenized isotropic cloak and the standard deviation ratio of a 50$\times$5 discretized anisotropic cloak (50 layers in the radial direction and 5 layers in the azimuthal direction), a 50$\times$10 discretized anisotropic cloak and a 50$\times$20 discretized anisotropic cloak.
As we can see, we achieve a better efficiency with the homogenized isotropic cloak than with the discretized anisotropic cloaks, which results from the asymptotic behaviour of $\kappa'_{\theta\theta}$. Thus even compared to the efficiency of the continuous Pendry's cloak in Fig. \ref{stdratio_Pendry_capeshomog} (red), it seems that the homogenization process has fairly optimal performances as the discretized anisotropic Pendry's cloak has a much lower efficiency for the same number of layers. We conclude that achieving efficiencies comparable to that of Pendry's perfect cloak appears to depend merely on the number $N$ of layers used in the homogenized cloak. A finer mesh in the neighbourhood of the cloaks' inner boundary should also improve their efficiency since this would better fit the rapid change of material parameters.

\section*{Methods}
	\subsection*{Time-harmonic heat equation}
	\label{time_harmonic_heat_equation}
We are here interested in a heat source resulting from the absorption of a modulated laser beam, so that the Fourier transform of temperature is introduced as:
\begin{equation}
u({\bf x}, \omega) = \int_{-\infty}^{+\infty} T({\bf x}, t)e^{-j\omega t}\mathrm{dt}
\end{equation}
where $T$ is the temperature defined in our computational domain, $t$ the time variable, $\omega$ the angular frequency and $\bf x$ the space variable.
We then take the Fourier transform of the heat equation 
\begin{equation}
{\rm div}(\kappa\nabla T)-\rho c\frac{\partial T}{\partial t} = 0
\end{equation}
leading to the time-harmonic heat equation
\begin{equation}
{\rm div}(\kappa\nabla u) +j\omega\rho c u = 0
\label{harmonic_heat_equation}
\end{equation}
where $u$ is the Fourier transform of the temperature, $\kappa$ is the conductivity, $\rho$ the density, $c$ the heat capacity and $\omega$ the angular frequency of a periodic heat source. Note here that this equation is valid in two and three dimensions.
	\subsection*{Standard deviation of isotherms}
The Fourier transform of the temperature $u$ can be written like the following
\begin{equation}
u = |u|e^{j\Phi}
\end{equation}
with $\Phi$ depending on the position and the angular frequency $\omega$. Thus, we retrieve the $(x,y)$-coordinate of the $|u|$ curves using COMSOL Multiphysics\textregistered\;and perform the standard deviation of $|u|$ for every isotherms.
	\label{standard_deviation_isotherms}
	\subsection*{Homogenization process}
	\label{homogenization_process} 
Following the proposal of Greenleaf and coworkers \cite{greenleaf08} to apply the two-scale convergence method of G. Allaire \cite{allaire92} and G. Nguetseng \cite{nguetseng89} in the design of approximate multi-layered cloaks, we derive a set of new effective parameters for a multi-layered heat cloak. We consider an alternation of concentric layers of respective conductivities $\kappa_1$ and $\kappa_2$ and of respective heat capacities $\rho_1c_1$ and $\rho_2c_2$ so that the overall conductivity and heat capacities can be written 
\begin{equation}
\begin{split}
\kappa_\varepsilon (x) &= \kappa_1 (x)\mathds 1_{[0,1/2]}\left(\frac{r}{\varepsilon}\right) + \kappa_2 (x)\mathds 1_{[1/2,1]}\left(\frac{r}{\varepsilon}\right) \\
\rho_\varepsilon c_\varepsilon (x) &= \rho_1 c_1 (x)\mathds 1_{[0,1/2]}\left(\frac{r}{\varepsilon}\right) + \rho_2 c_2(x)\mathds 1_{[1/2,1]}\left(\frac{r}{\varepsilon}\right)
\end{split} 
\end{equation}
where $x$ is the position vector, $r = ||x||$ is the Euclidean norm of $x$ and $\mathds 1_I$ is the indicator function of the set $I$. The function $\kappa_\varepsilon$ is periodic on the set $Y = [0,1]$ and $\varepsilon$ represents its periodicity (clearly, the thinner the layers in the cloak, the smaller the positive parameter $\varepsilon$). 
In this set of concentric layers, the time-harmonic heat equation is written
\begin{equation}
{\rm div}\left(\kappa_\varepsilon\nabla u_\varepsilon\right) +j\omega\rho cu_\varepsilon = 0
\end{equation}
where $u_\varepsilon$ is the Fourier transform of the temperature and $\omega$ is the angular frequency of a periodic heat source.
It is possible to show that when $\varepsilon \rightarrow 0$, this alternation of materials behaves like an anisotropic inhomogeneous medium of conductivity
\begin{equation}
A^{hom}(x) = \begin{pmatrix}
\frac{1}{\int_Y\frac{1}{\kappa_0}\mathrm{dy}} & 0 \\
0 & \int_Y \kappa_0\mathrm{dy}
\end{pmatrix}
\end{equation}
and heat capacity
\begin{equation}
(\rho c)^{hom}(x) = \int_Y \rho_0c_0\mathrm{dy} = \frac{1}{2}(\rho_1c_1+\rho_2c_2)
\end{equation}
where $\kappa_0$ and $\rho_0c_0$ are the respective two-scale limits of $\kappa_\varepsilon$ and $\rho_\varepsilon c_\varepsilon$ when $\varepsilon \rightarrow 0$ and are written
\begin{equation}
\begin{split}
\kappa_0 (x,y) &= \kappa_1 (x)\mathds 1_{[0,1/2]}(y) + \kappa_2 (x)\mathds 1_{[1/2,1]}(y) \\
\rho_0c_0 (x,y) &= \rho_1c_1 (x)\mathds 1_{[0,1/2]}(y) + \rho_2c_2 (x)\mathds 1_{[1/2,1]}(y)
\end{split}
\end{equation}
The homogenized time-harmonic heat equation is then written
\begin{equation}
{\rm div}\left(A^{hom}(x)\nabla u(x)\right) +j\omega(\rho(x) c(x))^{hom}u(x) = 0
\end{equation}
where $u$ is the limit of $u_\varepsilon$ when $\varepsilon \rightarrow 0$. Then, we need to solve the system
\begin{equation}
\left\{\begin{array}{l l}
A^{hom} = \underline{\kappa}' \\
(\rho c)^{hom} = \rho c\det(\bf{J})
\end{array}\right.
\end{equation}
for our set of materials to work as our cloak.

\section*{Acknowledgements}
We acknowledge the support of the Direction G\'{e}n\'{e}rale de l'Armement (DGA) and funding from ANR through INPACT project.

\section*{Author Contributions Statement}
D.P., S.G. and C.A. wrote the main manuscript text. M.B. provided help for the homogenization process method. M.Z. prepared figure 2 (left and right). All authors reviewed the manuscript.

\section*{Additional Information}
{\bf Competing financial interests:} The authors declare no competing financial interests.

\begin{table}[h!]
\begin{center}
\begin{tabular}{|*{11}{c|}}
\hline
$r\times 10^{-4} (m)$ & 1.5-1.65 & 1.65-1.8 & 1.8-1.95 & 1.95-2.1 & 2.1-2.25 \\
\hline
$\frac{\kappa_{hom}}{\kappa}$ & 240.35 & 0.07 & 10.01 & 0.13 & 6.30  \\
\hline
$\frac{\rho c_{hom}}{\rho c}$ & 0.19 & 0.52 & 0.80 & 1.04 & 1.24  \\
\hline
$r\times 10^{-4} (m)$ & 2.25-2.4 & 2.4-2.55 & 2.55-2.70 & 2.70-2.85 & 2.85-3 \\
\hline
$\frac{\kappa_{hom}}{\kappa}$ & 0.18 & 4.88 & 0.23 & 4.11 & 0.26 \\
\hline
$\frac{\rho c_{hom}}{\rho c}$ & 1.42 & 1.58 & 1.71 & 1.84 & 1.95 \\
\hline
\end{tabular}
\caption{\label{table_10_layer} Conductivity and product $\rho c$ parameters used for the 10-layer homogenized cloak after averaging on each radial element.}
\end{center}
\end{table}
\end{document}